\documentstyle[epsfig]{aipproc}

\def\beq{\begin{equation}}
\def\beql#1{\begin{equation}\label{#1}}
\def\eeq{\end{equation}}
\def\half{\mbox{\small $\frac{1}{2}$}}
\def\TeV{\mbox{\rm{TeV}}}       
\def\GeV{\mbox{\rm{GeV}}}       
\def\MeV{\mbox{\rm{MeV}}}

\begin{document}
\title{DIS 96--97. Theory/Developments\footnote{
Plenary talk given at DIS 97, Chicago, April 1997}}

\author{Yuri L. Dokshitzer}
\address{INFN, Sezione di Milano,\\
via Celoria 16, Milan 20133, Italy \\
and \\
St. Petersburg Nuclear Physics Institute, \\
Gatchina, St. Petersburg 188350, Russia.}
\maketitle

\begin{abstract}
Recent developments in the QCD understanding of DIS are reviewed, as well as
theoretical attempts to accommodate the HERA anomalous $e^+p$ events.
\end{abstract}

\section*{Introduction}

It was in dark 70's when the USSR was intact 
and whales did not enjoy full scale protection.
Back then a prominent Soviet solid state physicist Sergey V. Maleev,
returning from summer vacation, brought to his home institution in Leningrad 
a ``Whaler's Book''. He had found it in a little village high up,
deep in the Pamir Mountains.
The book contained, among other informative things, a formula
for calculating the volume of the catch. It ran 
\beql{whales}
  V \>=\> \pi\, R^2\, L\>.
\eeq
After having defined the dimensions $R$ and $L$, the author 
continued: 
\begin{quote}
``Here $\pi$ is an empirical constant which, for {\em Greenland whales},
equals approximately 3.14''. 
\end{quote}
Since then ``$\pi$ for Greenland whales'' has become a proverb in our
department.

I recalled this true story when 
reading an email message from Jose Repond who stressed
that the organisers of DIS 97 are expecting 
a coverage of theoretical attempts to explain the ``HERA anomaly''.

Let me start by complaining about the splitting of the ``Theory'' subject
into ``Theory/Phenomenology'' and ``Theory/Developments''. 
Physical Theory is a theory of Physical Phenomena. Therefore theoretical
developments aim at a better understanding
of Phenomenology, either directly or indirectly (we are not talking 
``theoretical theory'' here). 
Given this definition, 
it is still worth specifying 
two types, or rather two stages, of a physical theory:
\begin{center}
 Phenomenological Theory  (PT)\quad 
and \quad 
Theoretical Phenomenology (TP).
\end{center}
The distinction between the two is clear:
\begin{itemize}
\item[PT] is characterised by relative equilibrium. It has an intrinsic 
 logic of self-deployment, internal beauty, and relates to Phenomenology
 on a ``predict-check out-feed back'' basis.
\item[TP] is rather turbulent. For an insider, it is about a painful search
 for new principles to accommodate some ``out-of-hand'' 
 piece of Phenomenology.  
 For an outsider it is an easy target 
 for ridicule, simply because
 the initial ideas
 and constructions are bound to look naive, cumbersome, unnatural.
\end{itemize}
 
 TP may mark a transition period to a new PT, or a breakthrough in an existing
 one. Remember an outburst of ideas (open colour, 3 GeV weak bosons, etc.)
 which followed discovery of $J/\psi$.
 At the same time, it very well may be (and rather so often is) 
 leading nowhere, and with no-one to blame.
  
 Here ``the stakes we are gambling are frighteningly high''.
 What makes TP so vulnerable is that it emerges and develops as 
 a Theory of a {\em single}\/ Phenomenon (see Eq.~(\ref{whales})).

We shall discuss both: recent advances in QCD as Phenomenological Theory, 
and an explosive Theoretical Phenomenology triggered 
by the recent HERA anomaly.

\section*{QCD}
What makes the problem of DIS at very small $x$-Bjorken so difficult 
is its similarity to high-energy hadron-hadron interactions.
The latter subject was intensively discussed in 1960-70 and won 
the reputation of being one of the  most difficult and controversial   
``almost solvable'' problems in elementary particle physics, 
which it remains nowadays.

With the advent of QCD some key phenomena of high energy scattering have 
become less mysterious, at a qualitative level: 
(almost) constant total cross section (vector gluons), (almost) finite
transverse momenta of hadrons in minimum bias events (asymptotic freedom), 
fast decaying hadron form factors (internal quark structure), and the like. 
However, we have not come any closer to the solution of the problem of 
high energy hadron scattering which is intrinsically a large-distance
phenomenon. 
QCD in its present naive (quark-gluon; perturbative) form
may provide some hints but cannot address such a problem quantitatively.

\vspace{4mm}

The $x$-dependence of parton densities in normal hadrons is not perturbatively
predictable. It never was, neither for moderate nor for small values of $x$.
Perturbative QCD took responsibility for the {\em scaling violation}\/
pattern, i.e. for $Q^2$-evolution of parton distributions. At the same time
it has to rely on a non-perturbative initial input containing full
(arbitrary) $x$-dependence which one determines phenomenologically from
hard processes at some momentum transfer scale $Q_0$. 

The BFKL resummation programme aims exactly at this blind spot, as it attempts 
to {\em derive}\/ from first principles the $x$-dependence of parton
distributions, specifically at very small $x$.
Is this a perturbative problem in the first place? 
The answer is, strictly speaking, no.

A convincing argument is an argument that the part presenting 
it finds convincing. 
Given this definition, it becomes clear why critical voices
that spoke at previous DIS workshops
of the little relevance of the ``BFKL Pomeron'' to the HERA $F_2$
remained unheard: the criticism relied on convincing arguments.
Now the position of advocates of the thesis ``interesting but irrelevant'' 
has strengthened, since they've got respectable publications to refer to. 

\vspace{4mm}

We start from the paper by A.H.~Mueller entitled:

\begin{center}
``Limitation on using the operator product  expansion \\ 
 at small values of $x$'',
\qquad\qquad [hep-ph/9612251].
\end{center}
Mueller 
has explicitly found the boundary function $x_0(Q^2)$ 
``such that for Bjorken-$x$ values below $x_0$ 
the operator product expansion breaks down with 
significant non-perturbative corrections occuring in the leading twist 
coefficient and anomalous dimension functions due to diffusion 
to small values of transverse momentum''.

The author considers first
forward high-energy scattering of two tiny-size hadrons (onia)
with characteristic radii $R\ll 1{\rm fm}$.
Onium-onium scattering can be treated perturbatively as a hard process with  
an intrinsic hard scale 
$k_0=R^{-1}\gg\Lambda_{\mbox{\scriptsize QCD}}$.
Such a treatment was developed by Mueller some years ago. 
It was based on describing multiplication of soft gluons in the Fock 
states of the colliding hadrons, followed by an interaction 
between two small-$x$ gluon-partons.

In the large-$N_c$ approximation this multiplication can be described in terms
of the probabilistic colour dipole picture.   
Dipole density, determining the interaction cross section, is governed by
an evolution equation 
that naturally emerges in the impact parameter space.
Solution of the latter proves to be equivalent to the BFKL dynamics.

The upper bound on the total interaction energy, above
which the problem becomes entirely non-perturbative, reads:
\beql{yudok:OPE}
  \ln \frac{s}{M^2} = Y \><\> Y^{\mbox{\scriptsize  OPE}}(k_0)
= \frac{\pi}{14N_c\zeta(3)b^2}\left(\frac1{\alpha_s(k_0)}\right)^{3},\quad 
b=\frac{11N_c}{12\pi}-\frac{n_f}{6\pi}.
\eeq
Numerically,
\beql{yudok:OPEnum}
     Y^{\mbox{\scriptsize  OPE}}(k_0) \>\approx\> 
     \left(\frac1{2\alpha_s(k_0)}\right)^3.  
\eeq
At the same time, the Leading-Log (BFKL) approximation comes into a conflict
with the $s$-channel unitarity when $Y=\ln s\equiv \ln 1/x$ reaches so
large a value $Y^{\mbox{\scriptsize UN}}$
that the log-enhanced amplitude hits the limit 
\beql{yudok:hit}
 \alpha_s^2\> x^{-\omega_0}\>\simeq 1\>, 
 \quad \omega_0=\frac{N_c\alpha_s}{\pi}\>4\ln2 \>, 
\eeq
with $\omega_0$ the famous ``BFKL intercept''.
Thus,
\beql{yudok:UN}
  Y^{\mbox{\scriptsize UN}}(k_0) \approx \frac{2}{\omega_0}
  \ln \frac1{\alpha_s(k_0)}
  \>\simeq\> \frac1{\alpha_s(k_0)}\>. 
\eeq
Choosing a sufficiently large $k_0$ scale, so as to ensure
$1 \ll  Y^{\mbox{\scriptsize  UN}}\ll Y^{\mbox{\scriptsize  OPE}}$,
one can use onium-onium scattering as a theoretical laboratory for 
solving the unitarity quest within the perturbative QCD framework.

Returning to the DIS environment, we feel much less 
encouraged. 

Here basically the same boundary (\ref{yudok:OPEnum}) applies.
However, dealing with DIS structure functions we have two essentially 
different ``sizes'' (hard scales) in the game: 
that of the virtual photon, $Q$, and the initial hardness scale
$Q_0$, from which one dares to start the perturbative description.  
Clearly, now $k_0$ stands for the {\em minimal}\/ of the two, that is, 
$k_0\equiv Q_0$.
As a result, a head-on perturbative attack on the $x$-dependence of 
DIS structure functions is allowed only for the values of $x$ that respect 
\beql{yudok:OPEDIS}
   \ln\frac{\bar{x}}{x} \><\>  Y^{\mbox{\scriptsize  OPE}} \>\approx\> 
     \left(\frac1{2\alpha_s(Q_0)}\right)^3.  
\eeq
(Here $\bar{x}\sim 0.1$, according to the Mueller's estimate.)

Choosing the popular 1--2 GeV for $Q_0$ ($\alpha_s > 0.3$) 
leaves little room for perturbative treatment of the small-$x$ region. 
(Calculating cubes, keep in mind that when (\ref{yudok:OPEDIS}) is reached
the uncontrollable non-perturbative physics has already taken over! 
Step down.)

Lifting $Q_0$ to, say, 5--10 GeV would help theoreticians but won't please 
HERA experimenters since it undermines availability of the small-$x$ region
($Q^2> 25$--$100\> \mbox{GeV}^2$).

Experimental verification of a sharp increase of parton-parton scattering 
cross sections with energy, as predicted by BFKL dynamics, 
remains a challenging task. But it seems wiser to look for keys
not under a lamppost but where you lost them. 
In the BFKL context this means leaving in peace $F_2$ 
and addressing instead Mueller-Navelet jets 
and the transverse energy flow in DIS,
high-$p_t$ jets with large rapidity gap at hadron colliders, etc.  
(onium-onium scattering would certainly do as well).

A rare thing is more damaging to empirical science than an improper name.
In perturbative QCD we may discuss BFKL approximation, BFKL equation, 
BFKL dynamics but should never talk ``BFKL or Hard Pomeron''. 
There are two reasons for that. First of all, the Pomeron 
(Gell-Mann's name for the Gribov vacuum pole or vacuum singularity)
is a reserved word. 
(Would you dare to use a variable RETURN in your Fortran code?)
More importantly, the very term ``Hard Pomeron''
is nonsensical. 
\begin{itemize}
\item {\bf Pomeron:} 
  a leading high-energy contribution to elastic hadron scattering
  amplitudes driven by the leading singularity in the $t$-channel 
  (complex) angular momentum $\omega$. 
\item{\bf Hard:}  
  determined by small distances, and therefore perturbatively controllable. 
\end{itemize}
These two word simply don't merge, since the position and the nature of the
leading singularity in $\omega$ is entirely off the books of perturbative QCD!

One may generate the BFKL evolution to look for increase in 
cross sections, but should not refer to the {\em power}\/ ansatz 
(\ref{yudok:hit}) 
as a QCD (PQCD) prediction.

To avoid confusion we'd better switch from a ``BFKL (Hard) Pomeron''
to a ``BFKL (Hard) Heron'', an abbreviation for the High Energy Regime 
of 
small, small distance 
QCD parton scattering cross sections, 
applicable in a limited range of $\ln 1/x$.

\vspace{4mm}

The non-perturbative nature of the ``BFKL intercept'' has been (convincingly)
demonstrated by G.~Camici and M.~Ciafaloni in
\begin{center}
 ``Model (In)dependent Features of the Hard Pomeron''\qquad  
 [hep-ph/9612235].
\end{center}
The strategy adopted in the paper was to look at sensitivity of the
leading singularity to the behaviour of the QCD coupling in the infrared,
outside the perturbative domain. 
The second C\&C paper
\begin{center}
 ``K-Factorization and Small-$x$ Anomalous Dimensions''\qquad  
 [hep-ph/9701303]
\end{center}
dealt with the $q\bar{q}$ subleading contribution to the BFKL evolution
``with particular emphasis on running coupling effects''. In agreement with
the Mueller result the authors find that the next-to leading BFKL equation
is consistent ``at leading twist level, provided the effective variable
$\alpha_s(t)\ln 1/x$ is not too large''.

With account of the running coupling, 
the BFKL equation for the $t$-channel partial wave $f(t)$ 
with the angular momentum $\omega$ can be written as
\beql{yudok:BFKL}
  f(t)= f_0(t) + \frac{N_c\alpha_s(t)}{\pi\,\omega} 
  \int dt'\, K(t,t')\,f(t')\>,
 \qquad t=\ln\frac{k^2}{\Lambda^2}\>,\>\> 
        t'=\ln\frac{k^{\prime 2}}{\Lambda^2}\>,
\eeq
with $k$, $k'$ transverse momenta of gluons in the BFKL ``ladder''.
In terms of the spectral representation for the kernel,
$$
  K(t,t')\>=\> \int\frac{d\gamma}{2\pi i}\> \chi(\gamma)\> 
\exp\left\{(\gamma-\half)(t-t')\right\}, 
$$
the spectral function $\chi(\gamma)=\chi(1-\gamma)$ has a minimum 
$\chi_{\min} = \chi(\half)$.
The critical point $\gamma=\half$ generates a singularity in $\omega$
one is looking for.

The integral equation (\ref{yudok:BFKL}) is collinear safe.
Therefore, in the vicinity of this critical point, 
its kernel can be expanded as 
$$
  K(t,t')\>=\> \chi(\half)\left(1+\half \frac{\chi''(\half)}{\chi(\half)}
 + \ldots \right) \delta(t-t')\>,
$$
thus reducing (\ref{yudok:BFKL}) to a differential second order equation
$$ 
  f(t)-f_0(t) \>=\> c\frac{\alpha_s(t)}{\omega}
  \left( 1 + a^2 \partial_t^2\right) f(t)\>,
$$ 
with $a$, $c$ some constants.
The {\em homogeneous}\/ equation  for $f(t)$ is a Schr\"odinger equation,
$$ 
  \hat{H}(t)f(t)=\epsilon(\omega)\,f(t)\>, \quad 
  \hat{H}(t) = \partial^2 + V(t)\>, 
$$ 
with the ``potential'' 
$$ 
     V(t) \>=\> \mbox{const}\> \frac{\omega}{\alpha_s(t)}\>.
$$ 
It is the lowest spectral value of $\epsilon$ of this one-dimensional 
quantum mechanical problem that determines the 
leading singularity of the partial wave in $\omega$ 
(the ``Pomeron intercept'' $\omega_0$). 

For large positive values of $t$ the ``potential'' is well defined and grows,
due to asymptotic freedom, like $\alpha_s(t)^{-1}\propto t$. 
This information however is insufficient for solving the eigenvalue problem,
the obstacle being our ignorance about the region of small positive 
and negative $t$ (confinement, non-perturbative domain; you name it).

If we decide to {\em freeze}\/ the coupling below some $t=t_0$, 
the Schr\"odinger equation acquires a continuum spectrum 
(similar to the $\alpha_s=$const case), and the leading singularity
remains a cut in the complex $\omega$-plane. 

If instead we decide to {\em switch off}\/ abruptly gluon interaction 
at some low scale, we get a potential well 
(with an infinitely sharp left wall). 
The spectrum becomes discrete, and the leading singularity 
is a pole (followed by an infinite set of subleading poles).

Playing around with the shape of an effective $\alpha_s$ in the origin 
you may change the position of the QCD Pomeron and even its nature.
An explicit solution of the Schr\"odinger problem constructed by Camici and
Ciafaloni embodies the Pomeron singularity into the reflection coefficient
for scattering in the well potential. They conclude that the ``nature and 
location [of the Pomeron are determined by] soft hadronic interactions''.

{\flushright{Dixi.}} 

\vspace{4mm}

A long-awaited analysis of the next-to-leading corrections to the BFKL
evolution was reported by V.S.~Fadin, M.I.~Kotsky and L.N.~Lipatov in
\begin{center}
 ``Gluon pair production in the quasi-multi-Regge kinematics''\qquad  
 [hep-ph/9604267].
\end{center} 

In the BFKL language an exchanged $t$-channel (reggeized) gluon is called
``Reggeon'' and a real $s$-channel gluon --- a ``Particle''. 
Beyond the leading $(\alpha_s\ln 1/x)^n$ approximation virtual 
corrections emerge to the PPR and RRP vertices, as well as to the gluon
Regge trajectory. These corrections were calculated by Fadin, Fiore,
Kotsky and Quartarolo in 1995--96.
Now an analysis of the most technically difficult contribution 
due to production of two gluons (Particles) with comparable energies 
has been finalised. 
Together with a $q\bar{q}$ production channel (Camici and Ciafaloni)
all the bits and pieces are ready.

According to the 10$^{\mbox{th}}$ Commandment of Theoretical Physics,
``Only when you can explain it in 3 minutes, have you got a result:
neither God nor Landau will listen to you any longer''. From this point
of view, the FKL paper is not yet a result, though it marks a major 
breakthrough in the subject. 
Pieces still have to be assembled and looked into, to verify explicitly how
the coupling in the BFKL equation runs, and how much the next-to-leading
correction affects the rate of energy growth. 

Various Leading Log Approximations (LLA) are known to overestimate particle
multiplication. 
Therefore damping of  the predicted energy growth by the NL effects 
is a safe bet.
Some specific {\em physically motivated} NL corrections, in particular those 
stemming from kinematical effects, energy conservation, etc.),
can be studied  numerically with a help of an appropriate model.

\vspace{4mm}

In the paper by G.~Bottazzi. G.~Marchesini, G.P.~Salam and M.~Scorletti 
\begin{center}
 ``Structure functions and angular ordering at small $x$''
 [hep-ph/9702418]
\end{center} 
a special emphasis is given to angular structure of the final state glue. 
Their evolution picture reduces to the standard BFKL one in the LLA, 
but embodies essential subleading effects originating from the angular 
ordering, the latter being a consequence of the QCD coherence.
Analytic and numerical analysis has shown that due to the angular ordering 
the characteristic function $\chi(\gamma)$ gets modified, 
$\gamma_{crit}>\half$, the diffusion is reduced 
(the tendency competing with the running $\alpha_s(k_\perp)$ 
that should drag the system towards smaller transverse
momenta), and the singular $x$-behaviour becomes softer.

An important thing is that the BMSS study sets a framework for a forthcoming
theoretical analysis of the final state structure. 
It is worth remembering that the standard evolution equations 
designed to predict {\em inclusive}\/ cross sections (DGLAP, BFKL)  
exploit inclusive cancellations between real and virtual
contributions from unresolved partons and therefore are not suited,
generally speaking, for describing the structure of final states.
To make a long (and old) story short, in hard processes 
the final states are dominated by those partons which 
{\em do not contribute}\/ to the hard cross section 
(soft accompanying radiation). 
Perturbative predictions for final particle multiplicities, spectra, 
correlations etc. are Double-Logarithmic while the evolution of 
the basic cross sections is a Single-Log problem:
$$
  F^{LLA}(x;Q^2,Q_0^2) 
 = \sum_n a_n(x)\,\left(\alpha_s\ln\frac{Q^2}{Q_0^2}\right)^n, 
 \qquad \>\> \mbox{DGLAP, Infrared safe};
$$
$$
  F^{LLA}(x;Q_1^2,Q_2^2) 
 = \sum_n b_n(Q_1^2/Q_2^2)\,\left(\alpha_s\ln\frac{1}{x}\right)^n, 
 \qquad \mbox{BFKL, Collinear safe}.
$$

Now that you are convinced that there have been 
major developments in the field of small-$x$ physics since Rome-96, 
let me mention that this subject was far from
the only object of theoretical curiosity and labour.
In particular, a spectacular number of theoretical papers was devoted to
diffractive and quasi-diffractive particle production. 
Apologising for not having covered this and other QCD related topics 
in this talk, let me mention just one (the ``most theoretical'') paper
of the diffraction pool.

\vspace{4mm}

We started our discussion from the paper from Columbia University.
The closing one comes from British Columbia. 
It was written by 
Canadian physicists 
I.~Halperin and A.~Zhitnitsky and entitled
\begin{center}
 ``Hard diffractive electroproduction, transverse momentum distribution
   and QCD vacuum structure'' \qquad
 [hep-ph/9612425].
\end{center} 
This study lends a hope, if not a firm support, 
to those aiming at QCD phenomenology of diffractive processes. 
Elastic QCD scattering amplitudes were long known 
to exhibit the perturbative asymptotic regime only at very
high momentum transfer, $Q^2=$50--100 GeV$^2$. 
The authors analysed 
diffractive electroproduction of longitudinally polarised 
$\rho$-mesons 
within the QCD sum rule technology  
based on non-perturbative ITEP/OPE vacuum condensates.  
They have found that, in a marked contrast with the notorious property
of elastic scattering,  $\rho$-production 
shows a surprisingly low onset of the asymptotics starting at
$Q^2\simeq$10 GeV$^2$.

On this optimistic note we leave Phenomenological Theory.

\section*{HERA anomaly}

We turn to an outburst of theoretical activity that followed the
anomalous HERA events.  
Some personal data are due: I am not a big fan of supersymmetry 
(``interesting but irrelevant''), good old QCD providing 
enough headaches, to worry too much about any {\em newer}\/ physics.

The decision not to play truant and to dive into the subject has come with the
recognition that a reasonable part of the audience will probably 
have the same information background that I had, namely
\begin{enumerate}
\item Something weird is going on in $e^+p$ at HERA at the highest $Q^2$. 
\item {\em Its}\/ mass is 200 GeV.
\item The probability of {\em it}\/ being a statistical fluctuation is 1\%.
\end{enumerate}
What follows is an outsider's view on theoretical attempts to accommodate
the HERA anomaly. 

\begin{figure}[h!] 
\centerline{\epsfig{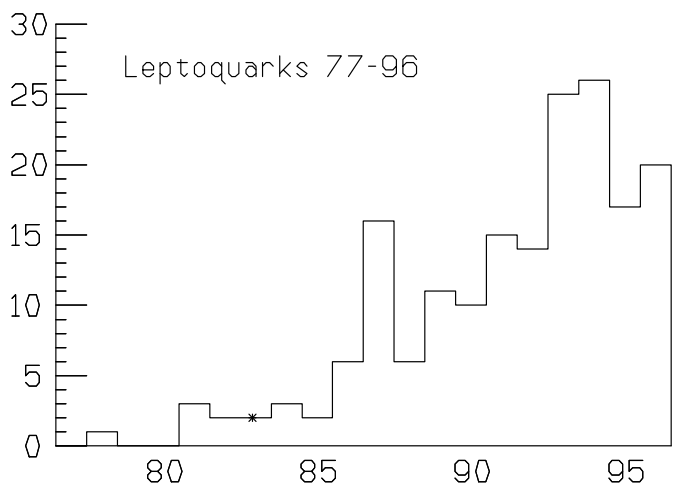}}
\vspace{10pt}
\caption{Statistics of leptoquark papers, 1977--1996.
The 1983 point marks the first title/abstract that responded 
to ``leptoquark'' and ``HERA'' simultaneously.}
\label{yudokst1}
\end{figure}

The theoretical discussion of leptoquarks (cards on the table!) 
has at least a 20-year (parabolic?) history displayed in Fig.~\ref{yudokst1}.
(The following statistics may fluctuate within a 15\% margin, 
depending on the database and search pattern chosen.)

\noindent
The past year's monthly statistics 
contains a prominent October bin (6 papers out of 20 total;
1\% probability for this being a statistical fluctuation, I gather).
%
%
%
Following 1+2 papers in January-February, the March--97 peak 
yields {\em more}\/ 
theoretical papers than the integrated 1996, with a similar
density in the first third of April.  
A list of 29 theoretical leptoquark/HERA-anomaly papers that appeared 
on the hep bulletin board before the Chicago Conference 
is assembled for your convenience at the end of this contribution.

Not that it was such a great joy to go through a couple of dozen of
preprints, many apparently having been written overnight.
At the end of the day, however, I found that forced journey quite
satisfactory. 
First of all, if you enjoy reading scientific papers written with passion, 
some papers on the list are worth your attention even if you are not an
expert in the field.
More importantly, I learned how much we actually know 
from experiment about Standard Model (SM) physics, 
the knowledge that leaves so
little room for going wild with our creative imagination.

The logic of the following short presentation was designed mainly 
with the help of the papers by 
 K.S.~Babu, C.~Kolda, J.~March-Russell and F.~Wilczek [9],
 J.~Kalinowski, R.~Rueckl, H.~Spiesberger and  P.M.~Zerwas [8],
 J. Bl\"umlein [7],
and
 G.~Altarelli, J.~Ellis, S.~Lola, G.F.~Giudice and  M.L.~Mangano [4].

\subsection{Contact interaction}
The starting point is to address the positron-quark interaction amplitude
$e^+q\to e^+q$ as a trouble-maker.
The standard procedure for parametrising unconventional physics consists of
writing down an effective point-like interaction Lagrangian involving some
new large momentum scale $\Lambda$. 
In our case there may be 8 a priori independent scales parametrising
16 options:
\beql{yudok:Leff}
  {\cal{L}} = \sum_{i,j=L,R} \frac{4\pi}{(\Lambda^q_{ij})^2}\> \eta^q_{ij}
  \> (\bar{e}_i\gamma_\mu e_i)\,(\bar{q}_j\gamma_\mu q_j)\>.
\eeq
Left/Right (L/R) options for the lepton and quark brackets give $2^2$,
$q=u/d$ another 2, and then for each of 8 scales there is a sign factor 
$\eta=\pm1$. 

Comments are due:
\begin{itemize}
\item
 A {\em scalar}\/ effective interaction is strongly vetoed by the 
 non-observation of helicity violating pion decays, 
 therefore only vector currents  in (\ref{yudok:Leff}).
\item
 Existing OPAL+CDF limits on $\Lambda_{ij}^q$ are typically 
 $\Lambda\sim 0.8 - 3$ TeV, depending on the channel.
\item
 The strongest constraints come from Atomic Parity Violation [9].
 For example, the Cesium nucleus ($Z=55$, $N\approx78$) was predicted to have
 the Standard Model total weak charge $Q_W^{SM}=-73.12\pm0.09$ which 
 value fits well with the observed $Q_W^{exp}=-71.04\pm1.81$. 
 Our contact interaction contributes to it as
 \beq
  \Delta Q_W = -2\left[\, C_{1u}(2Z+N) \>+\> C_{1d}(Z+2N)\,\right], 
 \eeq
 where
 $$
   C_{1q} = \frac{\sqrt{2}\pi}{G_F}\left\{ 
   \frac{\eta^q_{RL}}{(\Lambda_{RL}^q)^2}
  -\frac{\eta^q_{LL}}{(\Lambda_{LL}^q)^2}
  -\frac{\eta^q_{LR}}{(\Lambda_{LR}^q)^2}
  +\frac{\eta^q_{RR}}{(\Lambda_{RR}^q)^2} \right\}.
 $$
 A {\em single}\/ term in (\ref{yudok:Leff}) with $\Lambda<3$~TeV would produce
 $\Delta Q_W=\pm 20$, the resulting value being $10\sigma$ off.
\item
 The possibility of a {\em conspiracy}\/ between different helicity 
 contributions is not excluded. 
 Moreover, it has been argued by Ann~E.~Nelson [18] 
 to be quite natural, from inside the GUT scenario.
\item
 Actual fit to the $Q^2$-distribution of HERA events  
 based on the contact interaction was performed in [4]. It includes
 the less experimentally restricted scales $\Lambda^{-d}_{LR}$,  
 $\Lambda^{+d}_{RL}$,  $\Lambda^{+u}_{LR}$ and $\Lambda^{+u}_{RL}$
 and looks acceptable though not particularly spectacular.
\end{itemize}

\subsection{New exchange boson}
Now we are in a position to introduce more dynamics into the discussion
--- a new particle (boson) exchange as a model for the $e^+q\to e^+q$ blob.

We have three options to address:
\begin{itemize}
\item[$t\!\!$]-channel exchange, $e^+\to e^++X$, $q+X\to q$.
  This sort of $X$ has long been known as $Z'$. The combined D0+CDF 
  exclusion limit for
  direct production of such an animal in 
  $q+\bar{q}\to Z'\to e^++e^-$,~$\mu^+\mu^-$ is too high, $M_{Z'}>650$~GeV.
  Moreover, this mechanism would obviously provoke a similar signal 
  in $e^-p$ HERA sample as well, a reportedly unwanted feature.
\item[$u\!\!$]-channel. Here the interaction sequence is
 $e^+\to q+X$, $q+X\to e^+$. 
 In this case, however, the $e^-p$ DIS 
 would yield a much larger number of anomalous events.
 In this channel the interaction becomes resonant, and
 the crossing relation
$$
 \sigma(e^-p)\sim \frac{M_X}{\Gamma_X}\sigma(e^+p)
$$
would produce about $10^3$ $e^-p$ events at present statistics, 
given a typical $\Gamma\sim 10\,\MeV$.
\item[$s\!\!$]-channel production is the only viable option. 
 In this (as well as $u$-channel) case the exchanged
 boson obviously has both lepton and quark number, a {\em leptoquark}\/ (LQ).
\end{itemize}

\subsection{LQ quantum numbers}
The LQ is a colour triplet like a quark, bears weak $SU(2)$ isospin and may be
a scalar or a vector boson (leaving aside higher spins $J\!>\!1$ 
for the sake of renormalizability of a future Phenomenological Theory).

The leptoquark fermion number may be either $F\!=\!0$ or $F\!=\!2$.
The setup $F\!=\!0$, $J\!=\!0$ seems to be preferred.
\begin{itemize}
\item[$F\!=\!0$].
 The fermion number $F\!=\!2$ means merging the positron with 
 an {\em antiquark}\/ parton from the proton, LQ$=e^+\bar{q}$. 
 Such an option is ruled out.
 At large $x\sim 0.5$ where the game in being played, 
 the sea is suppressed by factor 50(200) with respect
 to the valence d(u) quarks.    
 Therefore, by the previous argument, 
 one would expect 
 {\em at least}\/ a few events (if not more)
 in the cross channel, $\overline{\mbox{LQ}}=e^-q$ in which the valence quarks 
 participate. Moreover, the sea suppression drives the value of the 
 coupling $\lambda$ quantifying the strength of the LQ$\ell\,q$ interaction 
 to dangerously large values to explain the observed signal.
\item[$J\!=\!0$].
 The situation with the LQ spin is not as obvious [7]. Still, the $J\!=\!1$ 
 option seems to balance on the brink of being excluded experimentally.
 LQs can be produced in pairs at the Tevatron via gluon exchange
 $q+\bar{q}\to g\to\mbox{LQ}+\overline{\mbox{LQ}}$.  
 The leptoquark production cross section has been limited 
 by D0 as $\sigma< 0.4$~pb.
 This translates into the upper limit for the vector LQ mass, 
 $M_{LQ}^{J=1}>240(215)\,\GeV$. A weaker restriction 215 emerges when the
 {\em maximal conspiracy}\/ between production via 
 the normal QCD-charge 
 and a potential anomalous colour coupling is allowed [9].
 In [4] the Tevatron production cross section of the {\em scalar}\/ LQ was
 estimated as $\sigma\simeq0.2$~pb, with $J\!=\!1$ overshooting by a factor 10.
\end{itemize}

This does not mean that the situation with $J\!=\!0$ LQ is rosy. 
The scalar leptoquark is also (and hopefully will remain) 
under severe pressure from the Tevatron. 
The combined CDF+D0 exclusion limit for the LQ mass is
$$
  M_{LQ} \><\> 190\>(143)\> \GeV\,, \qquad 
  \mbox{for}\>\> B(\mbox{LQ}\to e^++q)=1\>(\half)\>.
$$  
This brings us to an intriguing subject of  decay channels other
than $e^++q$.
 
A neutrino does not look easy to accommodate. An up-quark scenario 
$e^++u\to\mbox{LQ}$ calls for a weird object (system) with electric charge
$+5/3$ to accompany $\bar{\nu}_e$ substituted for $e^+$ in the final state. 
The $e^++d\to\mbox{LQ}$ channel looks fine, at first sight, since one
might expect a LQ decaying into $\bar{\nu}_e+u$.
However, by examining the $SU(2)$ doublet structure of the coupling 
one observes that the LQ must couple to both left and right leptons:
$e^+_L\,d_L$ combination to produce, $(\bar{\nu})_R\,u_R$ to decay.
This pattern, once again, is forbidden by pion decays.

It is worth noticing that an apparent difficulty with the neutrino may be 
good news, since no clear signal was reported in the Charged Current (CC) DIS.

\subsection{Special offer: squark}
An $R$-parity violating SUSY object --- squark --- as a LQ candidate seems to 
offer a natural qualitative explanation for some of the above-mentioned
peculiarities and troubles. Namely, it is ready to explain an absence of signal
in $e^-p$ and CC-$e^+p$ collisions and has room for decreasing the branching
into $e^+q$. 

$R=(-1)^{3B+L+2J}$ is +1 for normal particles (quarks, leptons, 
electroweak bosons) and $-1$ for their SUSY-partners. 
Within the $R$-parity-respecting scenario 
sparticles are produced in pairs.
Lifting off this requirement allows an $R$-parity-violating coupling of
a squark to a lepton and a quark, the one we are looking for.
The first thing to mention is that normal $R$-conserving SUSY decays of such
an object can be employed to push down the $e^+q$ branching, thus  
softening the Tevatron mass boundary.
 
Eight distinctive final state signatures for squark production 
were found and listed in [22]. 
Depending on the nature of the lightest {\em neutralino}\/  
(one among {\em photino}\/ $\tilde{\gamma}$, 
{\em zino}\/ $\tilde{Z}$ and two {\em Higgsinos} $\tilde{H}_0$; 
what a language!) they involve high-$P_\perp$ $e^+/e^-$ and/or missing
$P_\perp$ and different numbers of jets.

Minimal SUSY extension of the SM can be supplied with an abundance of 
$R$-violating structures:
\beql{yudok:susyR}
W_{\bar{R}} = \lambda_{ijk}\,L_iL_je_k^c 
             +\lambda'_{ijk}\,L_iQ_jd_k^c
             +\lambda''_{ijk}\,u_i^cd_j^cd_k^c\>.
\eeq
Capital $L$ and $Q$ stand for the left $SU(2)$-doublet lepton and quark fields.
Little $e$, $d$ and $u$ are right lepton and quark (up/down) singlets; 
superscript $c$ marks a conjugated field (antiparticle). 
Finally, $i,j,k$ are flavour generation indices, running from 1 to 3. 
Each field in (\ref{yudok:susyR}) is a {\em superfield}. This means that each
letter may be taken to represent either a normal particle or its super-partner.
In particular, the second term of this Lagrangian contains the combination
$\lambda'_{111}\>e_L^-\,\tilde{u}_L\, d_R^c$, the one responsible for 
the process $e^+_Rd_R\to \tilde{u}_L$. 
Similarly, the couplings $\lambda'_{121}$ and $\lambda'_{131}$ would produce
in $e^+d$ collisions $\tilde{c}$ and $\tilde{t}$ squarks respectively.
Thus we are talking about scalar SUSY-partners of the upper quark fields as
potential leptoquarks produced in positron-$d$-quark collisions.

Good old physics interferes again to veto the first generation scenario
involving $\tilde{u}$. 
Normal SUSY interaction may virtually transfer a $d$-quark into 
a spair $d\to\tilde{d}+\tilde{g}$.  
The new $R$-violating coupling $\lambda'_{111}$
then causes squark decay $\tilde{d}\to u+e^-$. 
This opens up a possibility of neutrinoless double $\beta$-decay
(put two neutrons close to each other and let their internal $d$-quarks 
exchange virtual gluino).

The HERA event rate implies $\lambda'>0.03$. At the same time, an 
experimental limit on $0\nu$ 2$\beta$-decay translates into a much smaller
value
$$
  \left|\lambda'_{111}\right| \><\> 7\cdot 10^{-3} 
\left(\frac{m_{\tilde{q}}}{200\,\GeV}\right)^2 
\left(\frac{m_{\tilde{g}}}{1\,\TeV}\right)^{\half}.
$$

The $\tilde{c}$ and $\tilde{t}$ options remain open. What is interesting 
(and may become a crucial turning point for upgrading the leptoquark TP 
into a PT, see above) is that each of the two scenarios 
predicts new discoveries ``around the corner'' [4]. 
Thus, the $\tilde{c}$-scenario provokes FCNC-forbidden decay 
$K^+\to\pi^+\nu\bar{\nu}$ not far below the present experimental limit, 
new effects in $D^0\leftrightarrow\bar{D}^0$ mixing,
$D^0\to e^+e^-$ and $D^+\to \pi^+e^+e^-$ decays. The $\tilde{t}$-scenario,
in turn, may cause a potentially large contribution to the $\rho$-parameter.

Under scrutiny are FCNC, CC-universality, neutrino masses, rare decays,
mixing, etc. 
The first impression is somewhat worrying. 
Experimental information about the 2$^{nd}$ and the 3$^{rd}$ generations,
though not as extensive as for the 1$^{st}$, severely restricts off-diagonal
{\em products}\/ of $\lambda'$- couplings.  In particular,
$\lambda'_{i12}\lambda'_{i21}$ ($\Delta M_K$ and $\epsilon_K$),
$\lambda'_{i13}\lambda'_{i31}$ ($\Delta M_B$),
$\lambda'_{11k}\lambda'_{21k}$ and 
$\lambda'_{1j1}\lambda'_{2j1}$ ($\mu {\rm Ti}\to e {\rm Ti}$),
$\lambda'_{1j1}\lambda'_{2j2}$ ($K_L\to \mu e$).
The list can be continued.
It looks as if we must have only one coupling to explain the HERA
anomalous events, and forbid all the others.
The question of ``naturalness'' comes to mind.
 
The $R$-violating Lagrangian (\ref{yudok:susyR}) possesses 
another conceptual problem. It contains, together with the $\lambda'$-sector, 
$\lambda$- and $\lambda''$-terms that violate FCNC and cause proton decay, 
respectively.
This problem was addressed by R.~Barbieri, Z.~Berezhiani and A.~Strumia 
in [28]. They argue that GUT initial conditions may
provide peace and calm down here by suppressing the unwanted couplings,
in a long course of evolution from the GUT scale,
in a more or less natural way.

Also under focus are virtual LQ effects in $e^+e^-$ collisions
(typically at the 1\% level, [8]),
prospects for LEP-2 and NLC studies [6], [16], [21], 
and an $e^+D$ option at HERA.
Results of the first calculations of QCD corrections to the Born
LQ production amplitudes have also been reported [19], [24].  

Highly non-orthodox solutions of the HERA puzzle were suggested by 
Stephen~Adler ($SU(4)$-preons) [1] and Alan~White (Sextet Quark Model) [27]. 
I am not in a position to provide you with a constructive criticism of these
ideas or to praise them. Nevertheless, a word of caution is due.
There is a strong chance that intrinsic ties between the electroweak and QCD
sectors of the SM are much deeper that we use to think they are. 
(Remember a mysterious pion, a point-like Goldstone from electroweak
point of view, and a loosely bound $q\bar{q}$ pair for QCD.)
Therefore, watch out and be ready for surprises.

Let me finish this non-expert review with a citation which has unintentionally
demonstrated how rapidly our field is actually developing:  
``... possible backgrounds come from the production and decay of top quark
pairs...'' [4]. It did not take long for the $t$-quark to develop into
a ``background'' for some newer and ever more exciting phenomena. 

The grip of the Standard Model remains tight.  
New physics? May be, it is about time. 
In a near future we will learn whether the
DIS-97 Chicago assembly marked a peak of a historic, or rather hysteric,  
period in elementary particle physics. 

\vspace{4 mm}
I am indebted to Gavin Salam for invaluable help.

\newpage
{\large\bf Anomaly related papers on the hep bulletin board 1997}

\noindent
\begin{itemize}
\item[28.]
   The high-$Q^2$ Hera anomaly and supersymmetric unification, \\
      Riccardo Barbieri, Zurab Berezhiani, Alessandro Strumia 
\hfill [04275]
\item[27.]
  Electroweak-Scale Excess Cross-Sections in the Sextet Quark \\
            ``Standard Model'',      Alan R. White 
\hfill [04248]
\item[26.]  $e^+ e^-$ Annihilation into Hadrons at LEP2 in the Presence 
     of \\
    the Anomalous DESY Positron-Jet Event Phenomenon, \\
     S. Jadach, B. F. L. Ward, Z. Was 
\hfill [04241]
\item[25.]  Stops in R-parity Breaking Model for High-$Q^2$ Events at HERA,\\
     T. Kon, T. Kobayashi 
\hfill [04221]
\item[24.]  QCD aspects of leptoquark production at HERA,\\
     C. Friberg, E. Norrbin, T. Sj\"ostrand 
\hfill  [04214]
\item[23.]  HERA high $Q^2$ events as indications of excited leptons 
     with \\
weak isotopic spin 3/2,
     B. A. Arbuzov 
\hfill [03460]
\item[22.]  R-Parity Violating Supersymmetry at HERA,\\
     Herbi Dreiner, Emanuelle Perez, Yves Sirois 
\hfill  [03444]
\item[21.]  Supersymmetry with R-Parity Breaking: Contact Interactions \\
     and Resonance Formation in Leptonic Processes at LEP2, \\
     J. Kalinowski, R. Rueckl, H. Spiesberger, P. M. Zerwas 
\hfill [03436]
\item[20.]  Formation and Decay of Scalar Leptoquarks/Squarks 
     in ep \\ 
collisions,
     T. Plehn, H. Spiesberger, M. Spira, P. M. Zerwas 
\hfill  [03433] 
\item[19.]  QCD Corrections and the Leptoquark Interpretation 
            of the \\
HERA High $Q^2$ Events,
     Z. Kunszt, W. J. Stirling  
\hfill [03427] 
\item[18.]  Contact Terms, Compositeness, and Atomic Parity Violation,\\
     Ann E. Nelson 
\hfill [03379]
\item[17.]  Four-Fermion Effective Interactions and Recent Data at HERA,\\
     Nicola Di Bartolomeo, Marco Fabbrichesi 
\hfill [03375]
\item[16.]  Leptoquark production at LEP2,
     Costas G. Papadopoulos 
\hfill [03372] 
\item[15.]  Like-Sign Dileptons at the Fermilab Tevatron Revisited 
            in \\
the Light of the HERA High-$Q^2$ Anomaly, \\
     D. Choudhury, S. Raychaudhuri  
\hfill [03369]
\item[14.]  Bounds on Contact Interactions from LEP1 Data and 
     the \\
High-$Q^2$ HERA Events,
     M. C. Gonzalez-Garcia, S. F. Novaes 
\hfill [03346] 
\item[13.]  Constraints on Leptoquark Masses and Couplings from 
     Rare \\
Processes and Unification,
     G. K. Leontaris, J. D. Vergados 
\hfill [03338]
\item[12.]  Much Ado About Leptoquarks: A Comprehensive Analysis, \\
     JoAnne L. Hewett, Thomas G. Rizzo 
\hfill [03337] 
\item[11.]  Removing flavor changing neutral interactions from leptoquark \\
 exchange, 
     M. Suzuki 
\hfill [03316]\
\item[10.]  Contact Interactions and high-$Q^2$ events in $e^+ p$ collisions 
     at HERA, \\
     V. Barger, Kingman Cheung, K. Hagiwara, D. Zeppenfeld 
\hfill [03311] 
\item[9.]  Comments on the high-$Q^2$ HERA anomaly \\
     K.S. Babu, Christopher Kolda, John March-Russell, \\
Frank Wilczek 
\hfill [03299] 
\item[8.]  Leptoquark/Squark Interpretation of HERA Events: 
     Virtual \\
 Effects in e+e-  Annihilation to Hadrons, \\
     J. Kalinowski, R. Rueckl, H. Spiesberger, P. M. Zerwas 
\hfill [03288]
 \item[7.]  On the Expectations for Leptoquarks in the Mass Range \\
 of O(200 GeV), 
     J. Bl\"umlein 
\hfill [03287] 
 \item[6.]  What Can We Learn About Leptoquarks At LEP200? \\
     Michael A. Doncheski, Stephen Godfrey 
\hfill [03285]  
 \item[5.]  High Q2-Anomaly at HERA and Supersymmetry,\\
     H. Dreiner, P. Morawitz 
\hfill [03279] 
 \item[4.]  Pursuing interpretations of the HERA large-Q2 data,\\
     G. Altarelli, J. Ellis, S. Lola, G. F. Giudice, M. L. Mangano 
\hfill [03276]
 \item[3.]  Rapidity Gap of Weakly Coupled Leptoquark Production in \\
ep Collider,
     T. K. Kuo, Taekoon Lee 
\hfill [03255] 
 \item[2.]  R-Parity Violation at HERA?\\
     Debajyoti Choudhury, Sreerup Raychaudhuri 
\hfill [02392] 
 \item[1.]  SU(4) Preonic Interpretation of the HERA Positron-Jet Events,\\
     Stephen L. Adler 
\hfill [02378] 
 \item[0.]  Scalar Leptoquark Pair Production at the CERN LHC: 
     Signal and \\ Backgrounds,
     B. Dion, M. de Montigny, L. Marleau, G. Simon 
\hfill [01285]
\end{itemize}

\end{document}